\documentclass[a4paper
]{article}
\usepackage{graphicx,amssymb,bm,latexsym}
\newcommand{\bea}{\begin{eqnarray}}
\newcommand{\ena}{\end{eqnarray}}
\newcommand{\vs}[1]{\vspace{#1 mm}}
\newcommand{\hs}[1]{\hspace{#1 mm}}
\renewcommand{\a}{\alpha}

\renewcommand{\d}{\delta}
\newcommand{\e}{\epsilon}
\newcommand{\s}{\sigma}

\newcommand{\la}{\lambda}
\newcommand{\pa}{\partial}

\newcommand{\nn}{\nonumber\\}
\newcommand{\p}[1]{(\ref{#1})}
\newcommand{\lan}{\langle}
\newcommand{\ran}{\rangle}

\begin{document}

\begin{titlepage}

\begin{flushright}
KU-TP 048
\end{flushright}

\vs{10}
\begin{center}
{\Large\bf Dark Energy and QCD Ghost}
\vs{15}

{\large
Nobuyoshi Ohta\footnote{ohtan@phys.kindai.ac.jp}} \\
\vs{10}
{\em Department of Physics, Kinki University,
Higashi-Osaka, Osaka 577-8502, Japan}

\vs{15}
{\bf Abstract}
\end{center}
\vs{5}

It has been suggested that the dark energy that explains the observed accelerating
expansion of the universe may arise due to the contribution to the vacuum
energy of the QCD ghost in a time-dependent background. The argument uses
a four-dimensional simplified model.
In this paper, we put the discussion in more realistic model keeping all components
of the QCD vector ghost and show that indeed QCD ghost produces dark energy proportional to
the Hubble parameter $H\Lambda_{QCD}^3$ ($\Lambda_{QCD}$ is the QCD mass scale)
which has the right magnitude $\sim (3\times 10^{-3}$ eV)$^4$.

\end{titlepage}
\newpage
\renewcommand{\thefootnote}{\arabic{footnote}}
\setcounter{page}{2}

\section{Introduction}

The recent cosmological observations have confirmed the existence of
the early inflationary epoch and the accelerated expansion of the present
universe~\cite{AC1}-\cite{AC3}. Observational result is consistent with
the picture that the universe has an unknown form of energy density, named
the dark energy, about 75\% of the total energy density.
The simplest possibility is the existence of the vacuum energy or cosmological
constant whose origin is yet to be identified.

Such vacuum energy is easily incorporated in the quantum field theory.
In the standard model of particle physics, we have Higgs field which produces
electroweak phase transition, which changes the value of the vacuum energy.
It has also been known that the vacuum fluctuations in quantum field theory naturally
induce such a vacuum energy, but the problem is how to control the size of it.
The contribution of quantum fluctuations in known fields up to 300 GeV, which is
about the highest energy at which the current theories have been verified, gives
a vacuum energy density of order (300 GeV)$^4$. This is vastly larger than
the observed dark energy density ($3\times 10^{-3}$ eV)$^4$ by a factor of order $10^{56}$.
Assuming the tree-level contribution is zero, it is a great challenge how to
understand the origin of this tiny energy density.

Recently a very interesting suggestion on the origin of a cosmological
constant is made, without introducing new degrees of freedom beyond
what are already known, with the cosmological constant of
just the right magnitude to give the observed expansion~\cite{U1}.
In this proposal, it is claimed that the cosmological constant arises from
the contribution of the ghost fields which are supposed to be present in the
low-energy effective theory of QCD~\cite{Witten,Veneziano,RST,NA,KO}.
The ghosts are required to exist for the resolution of the U(1) problem, but
are completely decoupled from the physical sector~\cite{KO}.
The above claim is that the ghosts are decoupled from the physical states and
make no contribution in the flat Minkowski space, but once they are in the
curved space or time-dependent background, the cancellation of their contribution
to the vacuum energy is off-set, leaving a small energy density $\rho \sim H \Lambda_{QCD}^3$,
where $H$ is the Hubble parameter and $\Lambda_{QCD}$ is the QCD
mass scale of order a hundred MeV. With $H\sim 10^{-33}$ eV, this gives
the right magnitude. This coincidence is remarkable and suggests that we
are on the right track.

However in this proposal, the authors discuss a four-dimensional model similar to the one
based on the Schwinger model (proposed by Kogut and Susskind~\cite{KS}),
keeping only the longitudinal and scalar components of the QCD ghost. These scalar fields have
positive and negative norms and cancel with each other, leaving no trace in the physical
subspace, but it is argued that they have small contribution to the vacuum energy in
the curved space  or time-dependent background.
(Similar system is used in a mechanism of supersymmetry breaking in Ref.~\cite{O1}.)
However it is known that the QCD ghost must be intrinsically vector field in order
for the U(1) problem to be consistently resolved within the framework of QCD~\cite{HKO}.
It is thus an interesting and important question to examine if the  mechanism
works even if we formulate the proposal keeping all the modes of the vector ghost.

In the next section, we briefly recapitulate how the U(1) problem is resolved
by the vector ghost following \cite{KO}, and also show how the ghost decouples from
the physical sector.
In sect.~3, following the discussions in Ref.~\cite{Z},
we discuss the mechanism of generating a tiny contribution to the vacuum energy
in the Rindler space as a typical example of the time-dependent
background~\cite{Unruh,BD}. We argue that due to the change of the definition of
the vacuum, we indeed obtain small contribution to the vacuum energy proportional
to the Hubble parameter.
We find that our result is factor 2 larger than the previous estimate.
Sect.~4 is devoted to the discussions and conclusions.

\section{Decoupling of the vector ghost in the Minkowski space}

We consider the low-energy effective Lagrangian~\cite{RST,NA,KO}
\bea
{\cal L} = \frac12(\pa_\mu S)^2 -\frac12 m_{NS}^2 S^2
+ \frac{1}{2F_S^2(m_S^2-m_{NS}^2)}(\pa_\mu K^\mu)^2 -\frac{1}{F_S} S\pa_\mu K^\mu,
\label{lag}
\ena
where $S$ is a flavor-singlet pseudoscalar field with the decay constant $F_S$,
$K^\mu$ is an axial vector ``field'', which in QCD corresponds to the gluonic current
\bea
K^\mu = 2N_f \frac{g^2}{16\pi^2} \e^{\mu\nu\la\s} A_\nu^a \Big(\pa_\la A_\s^a
+\frac13 g f^{abc}A_\la^b A_\s^c \Big),
\label{ano}
\ena
where $N_f$ is the number of flavors.

The Lagrangian~\p{lag} is invariant under the gauge transformation
\bea
K^\mu \to K^\mu + \e^{\mu\nu\la\s}\pa_\nu \Lambda_{\la\s},
\label{gi}
\ena
where $\Lambda_{\la\s}$ denotes an arbitrary antisymmetric tensor. In fact
this transformation reflects the color gauge invariance of the underlying QCD.
Under the QCD gauge transformation, the gluonic current transforms as~\cite{KO,HKO}
\bea
[Q_B, K^\mu]=2i N_f \frac{g^2}{16\pi^2} \e^{\mu\nu\la\s} \pa_\nu(C^a \pa_\la A_\s^a),
\ena
where $Q_B$ is the BRST charge and $C^a$ is the Faddeev-Popov ghost.

To quantize this system, we have to break the gauge invariance under~\p{gi}.
This can be done by adding the term
\bea
\frac{1}{4F_S^2(m_S^2-m_{NS}^2)\a} (\pa_\mu K_\nu-\pa_\nu K_\mu)^2,
\ena
where $\a$ is a gauge parameter. The simplest case is to choose $\a=1$.
One can then derive the Feynman rules as follows:
\bea
\mbox{$K_\mu$-propagator: } \frac{i\eta_{\mu\nu}}{k^2} F_S^2(m_S^2-m_{NS}^2), \quad
\mbox{$S$-propagator: } \frac{i}{k^2-m_{NS}^2}, \nn
\mbox{$S-K_\mu$-mixing: } \frac{1}{F_S}k_\mu. \hs{50}
\ena
It appears that the system~\p{lag} describes a scalar field $S$ with mass $m_{NS}$
and a massless vector. However it is not difficult to show that the mass of the scalar
$S$ gets shifted to $m_S$ due to the mixing of the scalar and vector modes.
This is the approach that Veneziano took in~\cite{Veneziano}.
Alternatively, in the general gauge, one can derive the two-point functions
\bea
T\lan K_\mu(x) K_\nu(y)\ran_M
&=& \int \frac{d^4 k}{(2\pi)^4}e^{-ik\cdot(x-y)}i \Big\{ \frac{k^2-m_{NS}^2}{k^2-m_S^2}
\frac{k_\mu k_\nu}{(k^2)^2} \nn
&& +\; \frac{\a}{k^2}\Big(\eta_{\mu\nu}-\frac{k_\mu k_\nu}{k^2}\Big) \Big\}
F_S^2(m_S^2-m_{NS}^2), \nn
T\lan S(x) S(y)\ran_M
&=& \int \frac{d^4 k}{(2\pi)^4}e^{-ik\cdot(x-y)} \frac{i}{k^2-m_{S}^2}, \\
T\lan S(x) K_\mu(y)\ran_M
&=& \int \frac{d^4 k}{(2\pi)^4}e^{-ik\cdot(x-y)}\frac{F_S(m_S^2-m_{NS}^2)}{k^2(k^2-m_{S}^2)},
\nonumber
\ena
where $T$ denotes the time-order and the subscript $M$ stand for the expectation value by
the Minkowski vacuum. We see that the two-point function of $S$ clearly shows
that it has the shifted mass $m_S$ instead of the original $m_{NS}$, confirming the
above claim. This is one of the consequences of the massless mode $K_\mu$
and gives a resolution of the $U(1)$ problem. What happens to the massless mode in the
system? Because we observe no massless field in the low-energy world,
it must decouple from the physical sector. The precise mechanism of this was not clear
in the approach of Ref.~\cite{Veneziano} but it was simply assumed that it decouples
because it is gauge-variant, as indicated above.
We now show how this can be achieved in the Minkowski space.

One can derive the field equations from the Lagrangian~\p{lag}:
\bea
&& \Box K_\mu - \pa_\mu \{ (1-\a)\pa^\nu K_\nu+F_S(m_S^2-m_{NS}^2)S \}=0, \nn
&& \Box S +m_{NS}^2 S+\frac{1}{F_S}\pa^\mu K_\mu=0.
\label{fe}
\ena
The first equation in ~\p{fe} tells us that $\Box K_\mu$ is expressed as a gradient
of a field, so we find
\bea
\Box(\pa_\mu K_\nu-\pa_\nu K_\mu)=0.
\ena
Hence we can consistently impose the subsidiary condition on the physical states:
\bea
(\pa_\mu K_\nu-\pa_\nu K_\mu)^{(+)}|\mbox{phys} \ran=0.
\label{ac}
\ena
In the gauge $\a=1$, we can write the mode expansion of the vector field.
In terms of the  canonically normalized field
$K'_\mu(x) \equiv F_S \sqrt{m_S^2-m_{NS}^2} K_\mu(x)$, it takes the form
\bea
K'_\mu(x) = \int \frac{d^3\bm k}{(2\pi)^{3/2}\sqrt{2k_0}} e_\mu^{(\la)}
[ e^{-ik\cdot(x-y)} a(\bm k,\la) + e^{ik\cdot(x-y)} a^\dagger (\bm k,\la)],
\ena
where $e_\mu^{(1)}=(0,\bm e^{(1)})$ and $e_\mu^{(2)}=(0,\bm e^{(2)})$ represent
the transverse modes, $e_\mu^{(3)}=(0,\bm e^{(3)})$ the longitudinal mode,
and $e_\mu^{(0)}=(1, \bm 0)$ the time component with
\bea
\bm e^{(3)} =\frac{\bm k}{|\bm k|}, ~~
\bm e^{(1)}\cdot\bm e^{(2)}=\bm e^{(1)}\cdot\bm e^{(3)}=\bm e^{(2)}\cdot\bm e^{(3)}=0.
\ena
Canonical quantization of the system then gives
\bea
[a(\bm k,\la), a^\dagger(\bm k',\la')] = \eta_{\la\la'} \d^3 (\bm k-\bm k'),
\ena
We see that the transverse modes
have the opposite sign to the usual gauge fields.
It is then easy to see that the condition~\p{ac} means that
the two transverse components $a(\bm k,1), a(\bm k,2)$ and the combination
$\frac{1}{\sqrt{2}}[a(\bm k,3)-a(\bm k,0)]$ should annihilate the physical state.
This forbids the states generated by the two transverse components and by
$\frac{1}{\sqrt{2}}[a^\dagger(\bm k,3)+a^\dagger(\bm k,0)]$. The remaining component
$\frac{1}{\sqrt{2}}[a^\dagger(\bm k,3)-a^\dagger(\bm k,0)]$ can only produce zero norm states,
so that all the components of $K_\mu$ are completely decoupled from the physical state.
Nevertheless it produces the physical effect of shifting the mass of the
singlet pseudoscalar $S$ and resolves the problem associated with the
$\eta'$ meson decay~\cite{KO}.
In the Kogut-Susskind model, similar subsidiary condition can be imposed~\cite{U1,O1}.

We are now going to see what effects this massless mode may produce
if our space is not just the Minkowski but curved space or time-dependent.

\section{Vector ghost in the Rindler space}

In this section, we consider the QCD vector ghost in the Rindler space.
Consider the Minkowski space
\bea
ds^2 = dt^2 - dx^2 - dy^2 - dz^2
\equiv d \bar u d\bar v - dx^2-dy^2,~~
\ena
where we have defined
\bea
\bar u=t-z,~~
\bar v= t+z.
\ena
Under the transformation
\bea
t=\frac{1}{a} e^{a\xi} \sinh a\eta,~~
z=\frac{1}{a} e^{a\xi} \cos a\eta,~~
(-\infty<\eta,\xi <\infty, a>0),
\label{rindler}
\ena
we obtain
\bea
ds^2 = e^{a(v-u)} du dv -dx^2-dy^2 =e^{2a\xi}(d\eta^2- d\xi^2)-dx^2-dy^2 ,\nn
\bar u = -\frac{1}{a} e^{a(\xi -\eta)} \equiv -\frac{1}{a} e^{-au},~~
\bar v = \frac{1}{a} e^{a(\xi +\eta)} \equiv \frac{1}{a} e^{av},~~~~~~~~
\ena
The Rindler coordinates $\eta$ and $\xi$ in~\p{rindler} describe only the quadrant part
$z>|t|$ called R. The opposite quadrant part L: $z<-|t|$ is described by changing
the signs in \p{rindler}. The rest of our Minkowski space are described by analytic
continuation of these coordinates~\cite{Unruh,BD}.

The wave function to be used in our massless vector field can be obtained from
the solutions for the scalar wave equation
\bea
\phi^{;\a}{}_{;\a}=0.
\ena
Explicitly this becomes in our coordinate system
\bea
[ e^{-2a\xi}(\pa_\eta^2-\pa_\xi^2)-\pa_x^2-\pa_y^2 ]\; \phi=0.
\ena
We denote by $^R u(\bm k)$ the wave function which asymptotes
\bea
^Ru(\bm k) =\left\{
\begin{tabular}{lcl}
$e^{-ik_0 u} e^{-i(k_1 x+k_2 y)}$ & \mbox{in} & \mbox {R} \\
0 & \mbox{in} & \mbox {L}
\end{tabular}
\right.
\ena
along the surface $v=-\infty, \bar v=0$, the past horizon of the Rindler coordinate.
Similarly the wave function in the L region is defined as
\bea
^Lu(\bm k) =\left\{
\begin{tabular}{lcl}
0 & \mbox{in} & \mbox {R} \\
$e^{ik_0 v} e^{-i(k_1 x+k_2 y)}$ & \mbox{in} & \mbox {L}
\end{tabular}
\right.
\ena

The positive-frequency Minkowski modes are characterized by the condition
that they are analytic and bounded in the lower half complex $\bar u$ plane on $\bar v=0$.
The combinations
\bea
\frac{1}{[2\sinh(\pi k_0/(2a))]^{1/2}} (e^{\pi k_0/(2a)}\; {}^Ru(\bm k)
+ e^{-\pi k_0/(2a)}\; {}^L u(-\bm k)^*), \\
\frac{1}{[2\sinh(\pi k_0/(2a))]^{1/2}}\; (e^{-\pi k_0/(2a)}\;{}^R u(-\bm k)^*
+ e^{\pi k_0/(2a)}\; {}^L u(\bm k)),
\ena
where $^L u(-\bm k)$ and $^R u(-\bm k)$ denote the wave function with minus momenta,
have precisely this property~\cite{Unruh,BD}, so we can use these modes
to express our Minkowski space field:
\bea
K'_\mu(x) &=& \int \frac{d^3\bm k}{(2\pi)^{3/2}\sqrt{2k_0}}
\frac{e_\mu^{(\la)}}{[2\sinh(\pi k_0/(2a))]^{1/2}}
[ (e^{\pi k_0/(2a)}\; {}^Ru(\bm k) + e^{-\pi k_0/(2a)}\;{}^L u(-\bm k)^*)a^{(1)}(\bm k,\la)\nn
&& + (e^{-\pi k_0/(2a)}\;{}^Ru(-\bm k)^* + e^{\pi k_0/(2a)}\;{}^L u(\bm k))a^{(2)}(\bm k,\la)
+ \mbox{h.c.} ],
\label{min}
\ena
and the Minkowski vacuum is defined as
\bea
a^{(i)}(\bm k,1)|0_M\ran =
a^{(i)}(\bm k,2)|0_M\ran =
[a^{(i)}(\bm k,3)-a^{(i)}(\bm k,0)]|0_M\ran =0,~~
(i=1,2).
\label{aux}
\ena

The field in the Rindler space is written as
\bea
K'_\mu(x) &=& \int \frac{d^3\bm k}{(2\pi)^{3/2}\sqrt{2k_0}}
e_\mu^{(\la)} [ {}^Lu(\bm k)\; b^{(1)}(\bm k,\la)+ {}^L u(\bm k)^*\; b^{(1)\dagger}(\bm k,\la)
 \nn
&&~~
+ {}^Ru(\bm k)\; b^{(2)}(\bm k,\la) + {}^R u(\bm k)^*\; b^{(2)\dagger} (\bm k,\la) ],
\label{rin}
\ena
Comparing \p{min} and \p{rin}, we see that
\bea
b^{(1)}(\bm k,\la) &=& \frac{1}{\sqrt{2\sinh\frac{\pi k_0}{a}}}
\Big[ e^{\pi k_0/(2a)} a^{(2)}(\bm k,\la)
 + e^{-\pi k_0/(2a)} a^{(1)\dagger}(-\bm k,\la) \Big], \nn
b^{(2)}(\bm k,\la) &=& \frac{1}{\sqrt{2\sinh\frac{\pi k_0}{a}}}
\Big[ e^{\pi k_0/(2a)} a^{(1)}(\bm k,\la)
+ e^{-\pi k_0/(2a)} a^{(2)\dagger}(-\bm k,\la) \Big].
\ena
The resulting energy for each mode is then given by
\bea
&& \lan 0_M|\int d^3 \bm k' \sum_{\la,\la'=0}^3 k_0 \eta_{\la\la'} b^{(1)\dagger}(\bm k,\la)
b^{(1)}(\bm k',\la') |0_M\ran \nn
&& =\lan 0_M|\int d^3 \bm k' \sum_{\la,\la'=0}^3 k_0 \eta_{\la\la'} b^{(2)\dagger}(\bm k,\la)
b^{(2)}(\bm k',\la') |0_M\ran
= \frac{4 k_0}{e^{2\pi k_0/a}-1}.
\ena
The vector field has four degrees of freedom and all of them are decoupled
in the flat Minkowski space. However, we find that they all contribute to
the vacuum energy in the Rindler space. This is somewhat surprising because
naively one would expect that the longitudinal and scalar modes would cancel.
Here instead of cancelling, they add up. This is the main result.
The result is factor 2 different from that in Ref.~\cite{U1},
but the order of magnitude is the same.

One may ask why such a difference occurred and whether the disagreement is resolved
if we consider the Faddeev-Popov (FP) ghost.
Actually the FP ghost was already discussed in Ref.~\cite{Z},
and the author concludes that the BRST charge does not annihilate the vacuum.
Under these circumstances it is not useful to use it to discuss the contribution.
However we should note that this system is abelian and the contribution can be discussed
perfectly well without such complication. (In fact it is argued that the ghost
does not make any contribution~\cite{Z}.)
Rather the difference comes not from this but from the treatment of the transverse modes.
It was assumed there that the transverse modes decouple and make no contribution, so
they were completely eliminated from the outset. This is true for the Minkowski space,
but the whole point of the discussions is that the result
and cancellation are modified when the system is considered in the time-dependent
background. So eliminating these modes from the start cannot be justified.
Thus it is expected that all four components contribute to the dark energy.

The contribution of high frequency modes is suppressed
by the factor $e^{-2\pi k_0/a}$ and the main contribution comes from $k_0\sim a$.
In the cosmological context, $a \sim H$ and hence $k_0 \sim H$, giving the vacuum
energy proportional to the Hubble parameter.
In the context of strongly interacting confining QCD with topological nontrivial sector,
this effect occurs only in the time direction and their wave function
in other space directions is expected to have the size of QCD energy scale.
As a result, this ghost gives the vacuum energy density $H\Lambda_{QCD}^3$ of
the right magnitude $\sim (3\times 10^{-3}$ eV)$^4$.
Thus this vacuum energy arises due to the mismatch between the vacuum energies computed
in slowly expanding universe and Minkowski space.

For the modes with $k_0\sim 1$ K $\sim 10^{-4}$ eV as in our present universe,
the contribution is suppressed by $\exp(-\frac{k_0}{H}) \sim \exp(-10^{29})$.
The deviation from Minkowski space starts only for modes with large wave
length $\la \sim a^{-1}$. Thus the effect is infrared in nature.
The local physics with $k_0 \gg a$ is not affected by the unphysical modes
with high accuracy.

\section{Discussions and conclusions}

In this paper, correcting the argument of Ref.~\cite{U1} in accordance with
QCD, we have first clarified the decoupling mechanism of the QCD vector ghost in the flat
Minkowski space, and then evaluated the contribution of the QCD vector ghost to the vacuum
energy density in the Rindler space as a typical example of time-dependent spacetime,
and found that it gives the vacuum energy proportional to the Hubble parameter.

This model has extremely interesting feature.
First of all, this does not assume new degrees of freedom only to produce
nonzero cosmological constant. Rather it is induced by the already existing field
just because the universe is expanding.
Secondly it gives the amazing result of
the cosmological constant of right magnitude without artificial fine tuning.
Note that the vacuum energy is not just a constant but depends on the Hubble parameter.

One may wonder that the unphysical modes or polarization of QED photon may
also contribute to the dark energy of the similar amount if the QCD vector ghost makes
such contribution. However, QED is weakly interacting theory unlike QCD and also
does not have nontrivial topological structure, and hence there is no restriction
on the wave function as in QCD. So the contribution to the energy density is very small
of order $H^4$ by dimensional reason and does not need to be considered~\cite{U1}.
However, see~\cite{JM} for alternative suggestion.

Other possible origin of such a ``vacuum energy'' was also suggested based on
the QCD condensate~\cite{S}, assuming that there is no contribution for
the flat Minkowski spacetime. Our mechanism is different in that such an assumption
is not necessary.

It has been suggested that the same ghost may also generate magnetic field
in an expanding universe~\cite{U2}. This is also discussed keeping only two
components of the QCD vector ghost.
It would be interesting to check if this mechanism makes sense with the vector ghost.
Another interesting question is to try to find if there is any other effects
to check the proposed mechanism. There are already some discussions on
this type of dark energy~\cite{Sol,ZC}.
These problems will be discussed elsewhere.

\section*{Acknowledgement}

I am grateful for valuable discussions with R.-G. Cai, H. Kodama and F. R. Urban.
This work was supported in part by the Grant-in-Aid for
Scientific Research Fund of the JSPS (C) No. 20540283, No. 21$\cdot$09225 and
(A) No. 22244030.


\end{document}